\documentclass[preprint,authoryear,12pt]{elsarticle}

\usepackage{graphics}
\usepackage{multirow}
\usepackage{amssymb}
\usepackage{lineno}
\usepackage{lscape}
\usepackage{color}

\journal{Icarus}

\begin{document}

\begin{frontmatter}

\title{Orbital clustering of Martian Trojans: An asteroid family in the inner solar system?}

\author[addr1]{Apostolos A.~Christou\corref{cor1}}
\ead{aac@arm.ac.uk}
\ead{Fax: +44 2837 527174}
\address[addr1]{Armagh Observatory, College Hill,
           Armagh BT61 9DG, Northern Ireland, UK}
\cortext[cor1]{Corresponding author}

\begin{abstract}
We report on the discovery of new Martian Trojans within the Minor Planet Center list of asteroids. Their orbital evolution over $10^{8}$ yr shows characteristic signatures of dynamical longevity
\citep{Scholl.et.al2005} while their average orbits resemble that of the largest known Martian Trojan, 5261 Eureka. The group forms a cluster within the region where the most stable Trojans should reside. 
Based on a combinatorial analysis and a comparison with the Jovian Trojan population, we argue that both this feature and the apparent paucity of km-sized Martian Trojans \citep{Trilling.et.al2006} as compared to expectations from earlier work \citep{TabachnikEvans1999} is not due to observational bias but instead a natural end result of the collisional comminution \citep{Jutzi.et.al2009} or, alternatively, the rotational fission \citep{Pravec.et.al2010} of a progenitor $\mbox{L}_{5}$ Trojan of Mars. Under the collisional scenario in particular, the new Martian Trojans are dynamically young, in agreement with our age estimate
of this ``cluster'' of $<$ 2 Gyr based on the earlier work of \citeauthor{Scholl.et.al2005}.       
This work highlights the Trojan regions of the Terrestrial planets as natural laboratories to study processes important for small body evolution in the solar system and provides the first direct evidence for an orbital cluster of asteroids close to the Earth.
\end{abstract}
   

\end{frontmatter}


\section{Introduction}
Trojan asteroids are objects confined by solar and planetary gravity to orbit the Sun $60^{\circ}$ ahead ($\mbox{L}_{4}$) or behind ($\mbox{L}_{5}$)  a planet's position along its orbit \citep{MurrayDermott1999}.
They hold a special status among asteroids. Due to their dynamical longevity \citep{Levison.et.al1997,Marzari.et.al2003a,Marzari.et.al2003b,Scholl.et.al2005}, they are thought to
be left over material from the formation and early evolution of our planetary system \citep{Shoemaker.et.al1989,MarzariScholl1998a,MarzariScholl1998b,Marzari.et.al2002,Morbidelli.et.al2005,NesvornyVokrouhlicky2009,LykawkaHorner2010}.
Mars is the only terrestrial planet currently known to host stable Trojans \citep{Scholl.et.al2005,Dvorak.et.al2012}. As of 21 June, 2012, only three long-lived objects appear in the Minor Planet Center (MPC) list of Martian Trojans (http://www.minorplanetcenter.net/iau/lists/MarsTrojans.html): 
5261 Eureka ($\mbox{L}_{5}$), (101429) 1998 $\mbox{VF}_{31}$ ($\mbox{L}_{5}$) and (121514) 1999 $\mbox{UJ}_{7}$ ($\mbox{L}_{4}$). Inclusion in this list requires that they be confirmed as Trojans by numerical integrations for at least $10^{5}$ yr (Gareth Williams, priv.~comm.). Several other objects are temporary co-orbitals of Mars 
\citep{Connors.et.al2005}, similar to Earth's transient co-orbitals \citep{Namouni.et.al1999,Christou2000a,MoraisMorbidelli2002,Brasser.et.al2004b,Wajer2010}.
The debate of the existence and origin of long-lived terrestrial planet Trojans is now coming to the fore \citep{Todd.et.al2012a,Todd.et.al2012b} due to expected results from the upcoming GAIA \citep{Mignard.et.al2007} and NEOSSSAT \citep{Laurin.et.al2008} missions as well as new deep, wide field surveys of the sky such as the Panoramic Survey Telescope \& Rapid Response System \cite[PanSTARRS;][]{Jedicke.et.al2007} and the Large Synoptic Survey Telescope \cite[LSST;][]{Jones.et.al2009}. A Trojan of the Earth was discovered recently \citep{Connors.et.al2011} and long-lived Trojans of our planet may exist \citep{Dvorak.et.al2012}. 

The expected number of Martian Trojans $\gtrsim 1$ km in diameter has been estimated to be $\lesssim 50$ \citep{TabachnikEvans1999}. Previous studies \citep{TabachnikEvans1999,TabachnikEvans2000a,Scholl.et.al2005} have shown that the most long-lived Trojan librators of Mars exist in particular regions of orbital element space. 
\citet[][hereafer referred to as SMT05]{Scholl.et.al2005} mapped out Trojan stability as a function of their proper elements, quantities that are constants of motion under planetary perturbations \citep{Milani1993,BeaugeRoig2001}. They found that the most stable objects are those with proper eccentricity $e_{p}<0.15$, 
proper inclination $13^{\circ} <I_{p}<28^{\circ}$ and proper libration amplitude $0^{\circ}<D<150^{\circ}$. Eureka and 1999 $\mbox{UJ}_{7}$ ($e_{p}\sim 0.05$, $I_{p}\sim 22^{\circ}$) occupy the projection of this stable island on the $e_{p}-I_{p}$ plane, whereas 1998 $\mbox{VF}_{31}$'s higher proper inclination ($I_{p}\sim30^{\circ}$) places it in a different 
region, separated from the stable island by the $\nu_{5}$ secular resonance. Both 1999 $\mbox{UJ}_{7}$ and 1998 $\mbox{VF}_{31}$ 
have proper libration amplitudes $\gtrsim 50^{\circ}$, unlike Eureka ($\sim 10^{\circ}$). Further, SMT05 found that long-lived Trojans 
(ie not the transitory types investigated by \citet{Connors.et.al2005}) belong to two groups: the ``fast diffusers'' (FD) 
that escape from Trojan libration within $10^{8}$ yr and the ``slow diffusers'' (SD)
that persist as Trojans of Mars for periods of time comparable to the age of the solar system. 

Here we report on a study of new candidate Martian Trojans discovered within the Minor Planet Center list of asteroids. The osculating orbital elements of these objects are similar to Eureka's, suggesting that they may also be long-lived. This work aims to determine, in the first instance, whether these are transitory, and if not, whether they belong to the SD 
or the FD group. Additionally, we wish to constrain the nature of their relationship, if any, to Eureka and the other known SD Trojans. 
In the next Section we describe the new Trojan candidates. In Sections~\ref{supp:int} \& \ref{supp:proper}
we establish their stability and relation to the other known Trojans.
In Section~\ref{supp:stats} we quantify the statistical significance of the observed orbital clustering while
in Section~\ref{supp:obs} we argue that it is likely not an artifact of observational bias. In Section~\ref{supp:jupiter} we 
compare the Martian and Jovian Trojan populations. In Section 8 we place 
our findings in context and discuss different scenarios for the origin and evolution of these objects 
to the present date. In the last Section we summarise our main conclusions and outline the work's implications
for small body research. 

\section{Search for new candidate Martian Trojans} 
\label{supp:cands}
We searched through the Minor Planet Center database\footnote{Available at http://www.minorplanetcenter.net/db\_search} and found 7 asteroids with similar osculating semimajor axis, 
eccentricity and inclination to Eureka. The specific criteria used were: $1.523\mbox{ AU}<a<1.5245\mbox{ AU}$, $e<0.15$ and $16^{\circ}<I<24^{\circ}$. Table~\ref{tab:allobj} shows
these eight objects and, in addition, the long-term stable Trojan with significantly higher inclination than Eureka, 1998 $\mbox{VF}_{31}$. We consider the six objects that are not mentioned in the MPC
list to be new candidate Martian Trojans.

Three of the new candidates, (311999) 2007 $\mbox{NS}_{2}$, 2011 $\mbox{SC}_{191}$ and 2011 $\mbox{UN}_{63}$ have been observed at multiple oppositions, hence their orbits
should be well determined. The remaining three, 2011 $\mbox{SL}_{25}$, 2011 $\mbox{SP}_{189}$ and 2011 $\mbox{UB}_{256}$ are only single-opposition discoveries, 
implying that their orbits are uncertain. Therefore we have not considered those further in this paper and focus our attention on the three
objects with well-determined orbits. 
2007 $\mbox{NS}_{2}$ was discovered by OAM Observatory, La Sagra (J75) and was subsequently precovered in 1998 LONEOS and LINEAR 
survey data (MPEC 2007-O09).
2011 $\mbox{SC}_{191}$ is a recovery, by the Mt Lemmon survey, of 2003 $\mbox{GX}_{20}$, an object initially discovered by the NEAT survey.
Similarly, 2011 $\mbox{UN}_{63}$ is a recovery, by the Mt Lemmon survey, of one of its own discoveries, 2009 $\mbox{SA}_{170}$. 

Apart from 5261 Eureka, (101429) 1998 $\mbox{VF}_{31}$ and (121514) 1999 $\mbox{UJ}_{7}$, the study by SMT05 also demonstrated long-term stability  for 2001 $\mbox{DH}_{47}$.
This object also satisfied the criteria of our search through the MPC database but does not appear in the Minor Planet Center list of Martian Trojans as do the other three objects \cite[see also][]{Todd.et.al2012b}. Interestingly, it has not yet been awarded a number despite having been a multi-opposition object since 2003. These ``anomalies'' cast its status as a stable Trojan of Mars and/or its orbit in doubt  and so it has not been included in Table~\ref{tab:allobj}. We have nevertheless considered it in the combinatorial analysis (Section~\ref{supp:stats}) insofar as to show that the main conclusions of the paper do not change if it is included.

\section{Establishing Trojan stability by long-term numerical integrations}
\label{supp:int}

To determine (i) whether they are, in fact, Trojans of Mars, (ii) their likely lifetime in Trojan libration, and (iii) if not transitory, whether they belong to either the FD or SD populations (SMT05), 
we resort to numerical integration of their orbits. The covariance ellipsoid for each of these three objects and Eureka was populated with 101 clones using the method described in 
\citet{Duddy.et.al2012}. These clones were then integrated for $10^{8}$ yr in the past  under a force model that included the gravity of the 8 major planets, the dwarf planet Ceres and the large Main Belt asteroid Vesta. 
The code utilises the second-order mixed variable symplectic (MVS) algorithm implemented within the MERCURY package \citep{Chambers1999}.
A model of the along-track component of the diurnal Yarkovsky acceleration \citep{Farinella.et.al1998} was also incorporated 
into MERCURY's user-defined force feature. In this model, the magnitude of the Yarkovsky acceleration $\alpha_{\rm Y}$ has the form

\begin{equation}  \label{eq:yarko}
\alpha_{\rm Y}=\frac{2}{\rho R}\frac{\epsilon \sigma T^{4}}{c}\frac{\Delta T}{T} \cos \zeta \\
\end{equation}

where $\rho$ is the bulk density, $R$ the object radius, $\epsilon$ is the surface thermal emissivity, $\sigma$ is the Stefan-Boltzmann 
constant, $T$ the surface temperature and $c$ the velocity of light, $\zeta$ the rotational axis obliquity. $\Delta T / T$ is the effective temperature difference around the surface of the body and depends on $\rho$ and $R$ as well as the surface albedo $A$, specific heat capacity $C$, thermal conductivity $K$ and rotational frequency $\nu$. The magnitude of the force depends on the heliocentric distance through the term $\Delta T / T$ which is a function of solar insolation. In the integrations, the heliocentric distance used in Eq.~\ref{eq:yarko} was kept fixed at 1.524 AU, the semimajor axis of the Martian orbit.

Two separate batches of $N=101$ clones per object were integrated. In the first batch the Yarkovsky force was switched off and the system is conservative.
In the second, 
each one of  $N$ randomly-generated clones were assigned a value of the Yarkovsky acceleration 
${\rm \alpha_{Y}}={\rm \alpha_{Y,max}} \left(2 i/N - 1\right)$, $i=0,1,\cdots, N$. The value of ${\rm \alpha_{Y,max}}$ was calculated from Eq.~\ref{eq:yarko} by setting $\zeta=0$, an assumed bulk density of 1 g $\mbox{cm}^{-3}$ and equal to the surface density,  
$K = 4\times 10^{-3}$ W $\mbox{m}^{-1}$ $\mbox{K}^{-1}$, $C=680$ J $\mbox{kg}^{-1} \mbox{K}^{-1}$,  $\epsilon=0.88$ and $A=0.12$. The resulting values of ${da/dt}_{max}$ $\left(=2 \alpha_{Y,max}/n\mbox{, n being the mean motion}\right)$ 
for the four objects are: 80 m $\mbox{yr}^{-1}$ (5261), 150 m $\mbox{yr}^{-1}$ (311999), 450 m $\mbox{yr}^{-1}$ (2011 $\mbox{SC}_{191}$) and 480 m $\mbox{yr}^{-1}$ (2011 $\mbox{UN}_{63}$).
 
A rotation period of 6 hr was used for Eureka \citep{Rivkin.et.al2003} and
the same value assumed for the other asteroids. For the radius we adopted the value of $0.65$ km for Eureka \citep{Trilling.et.al2007}. For the other objects, the radii were calculated from the absolute magnitude assuming an albedo of 0.5 using the MPC absolute magnitudes to diameter conversion tables\footnote{Available at http://www.minorplanetcenter.net/iau/lists/Sizes.html}. This is done to err on the side of caution and overestimate the actual Yarkovsky acceleration on the asteroids by adopting the smallest size that is consistent with their absolute magnitudes.

\begin{figure}
\vspace{-3cm}
\centering
\includegraphics[width=150mm,angle=0]{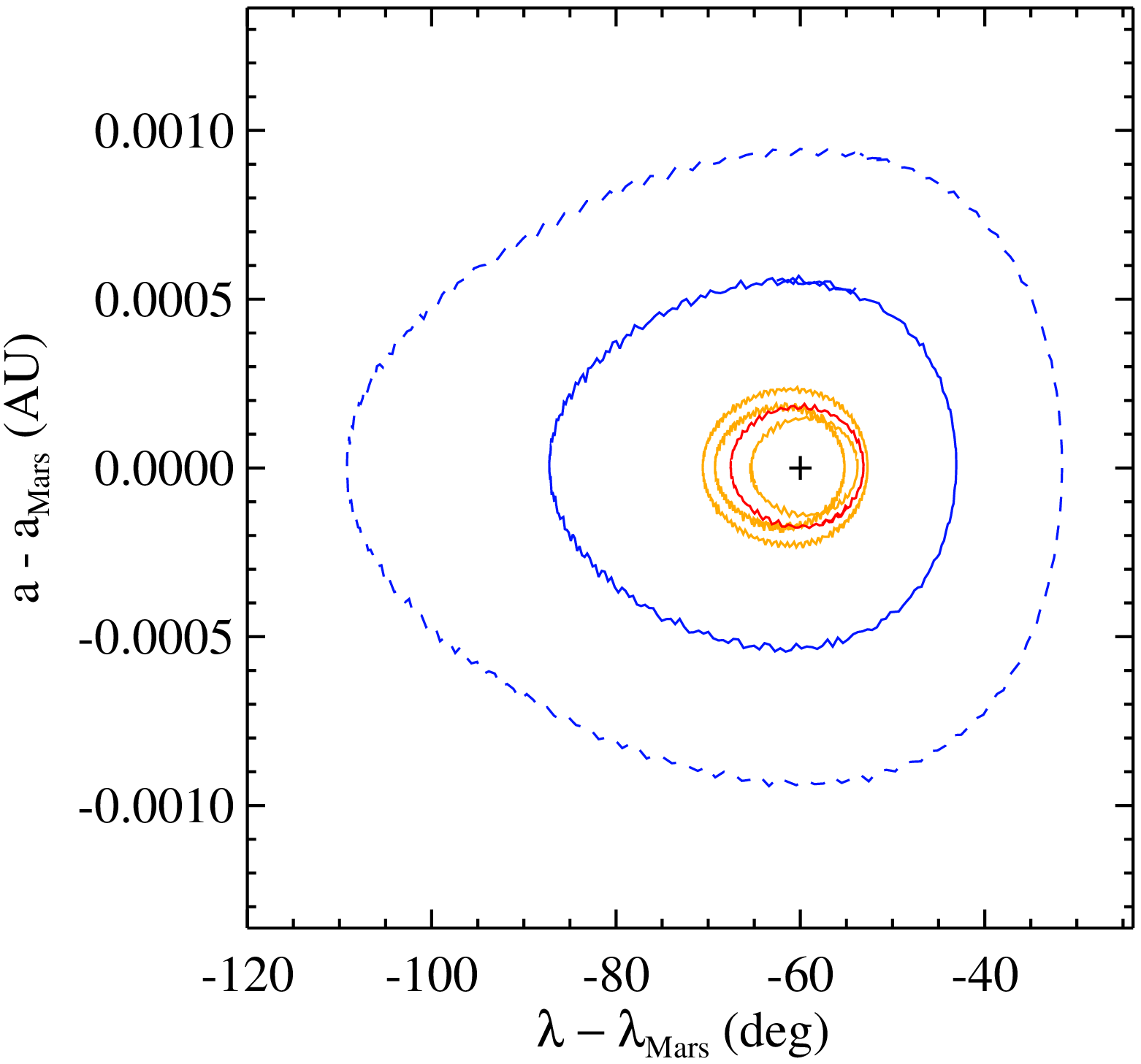}
\caption[Libration around the corresponding equilibrium point (plus sign) for all six long-lived Martian Trojans considered in this work.  
In order of decreasing libration amplitude, they are: (121519) 1999 $\mbox{UJ}_{7}$ ($\mbox{L}_{4}$; dashed blue curve), (101514) 1998 $\mbox{VF}_{31}$ ($\mbox{L}_{5}$; blue curve), (5261) Eureka ($\mbox{L}_{5}$; red curve); (301999) 2007 $\mbox{NS}_{2}$, 2011 $\mbox{SC}_{191}$ and 2011 $\mbox{UN}_{63}$ ($\mbox{L}_{5}$; amber curves).]{}
\label{fig:avsl}
\end{figure}

In the first instance, we observe that the orbital mean longitude $\lambda - \lambda_{\rm Mars}$ and semimajor axis $a - a_{\rm Mars}$ of the new objects relative to Mars librate around the location of the $\mbox{L}_{5}$ equilibrium point (Fig.~\ref{fig:avsl}). The intersection of the curves for the smaller amplitude cases is likely due to the significant orbital inclination of Martian Trojans which shifts the libration center \citep{NamouniMurray2000}.

\begin{figure}
\vspace{-3cm}
\centering
\includegraphics[height=4cm,angle=00]{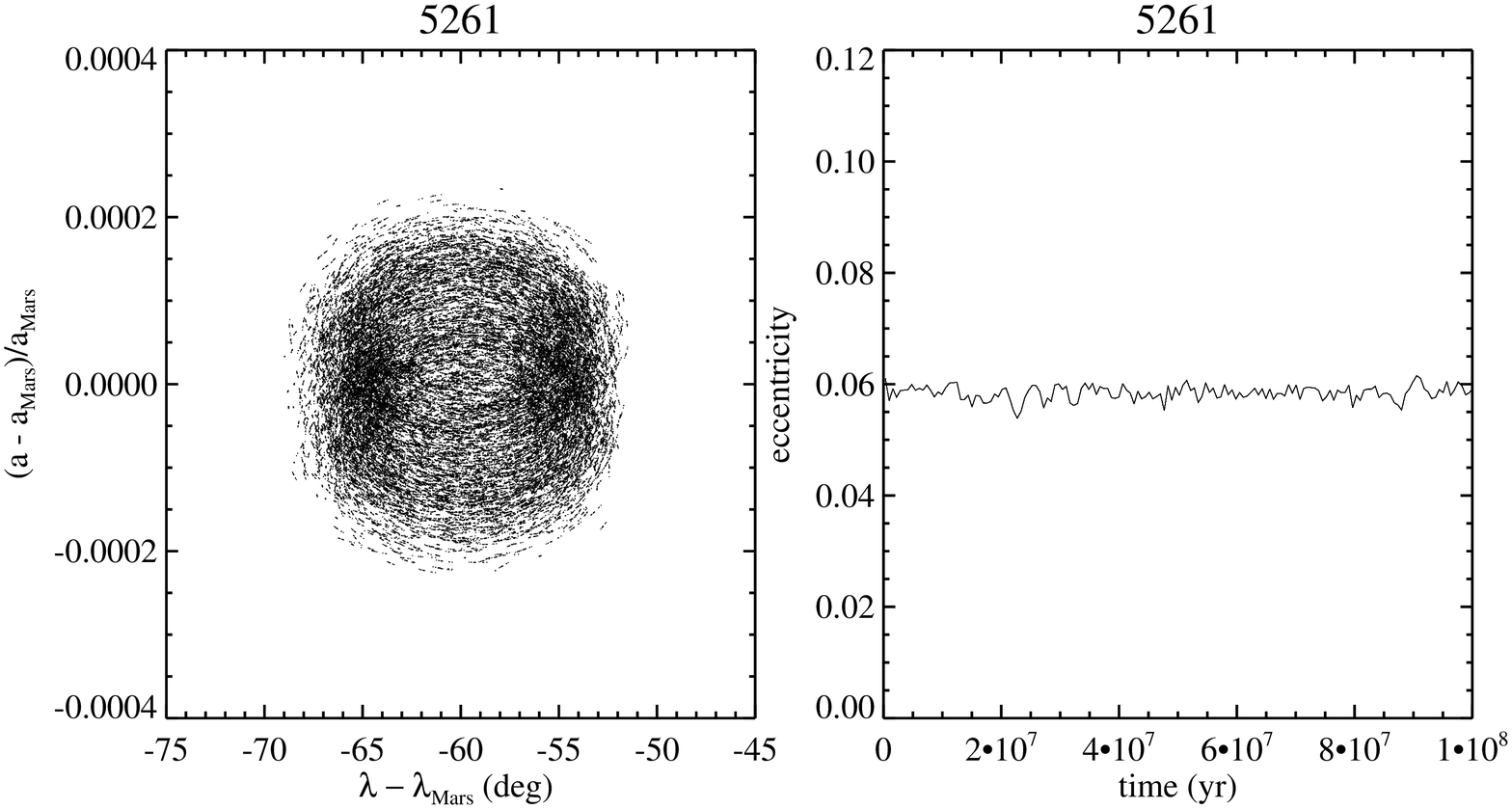}\\
\includegraphics[height=4cm,angle=00]{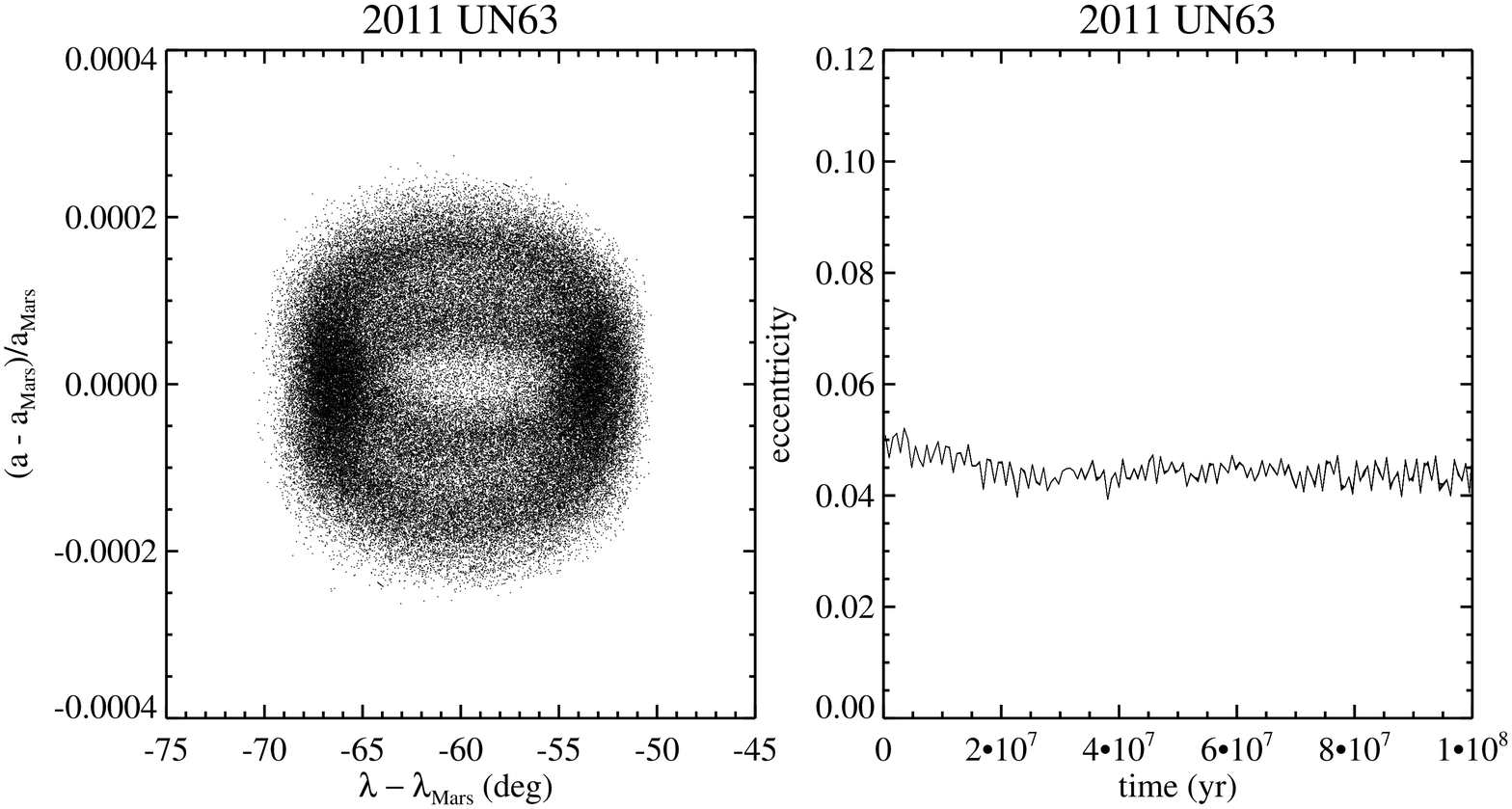}\\
\includegraphics[height=4cm,angle=00]{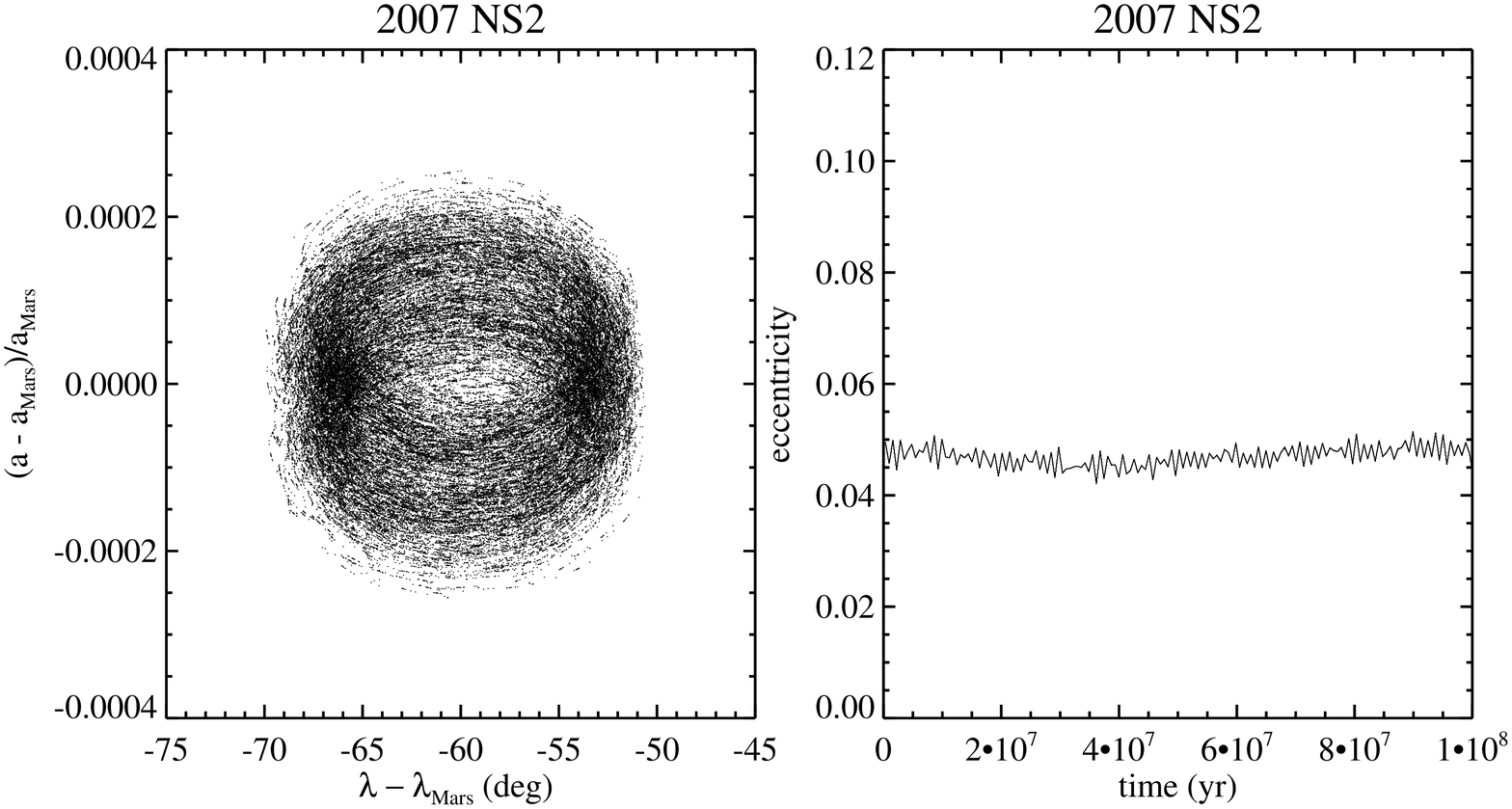}\\
\includegraphics[height=4cm,angle=00]{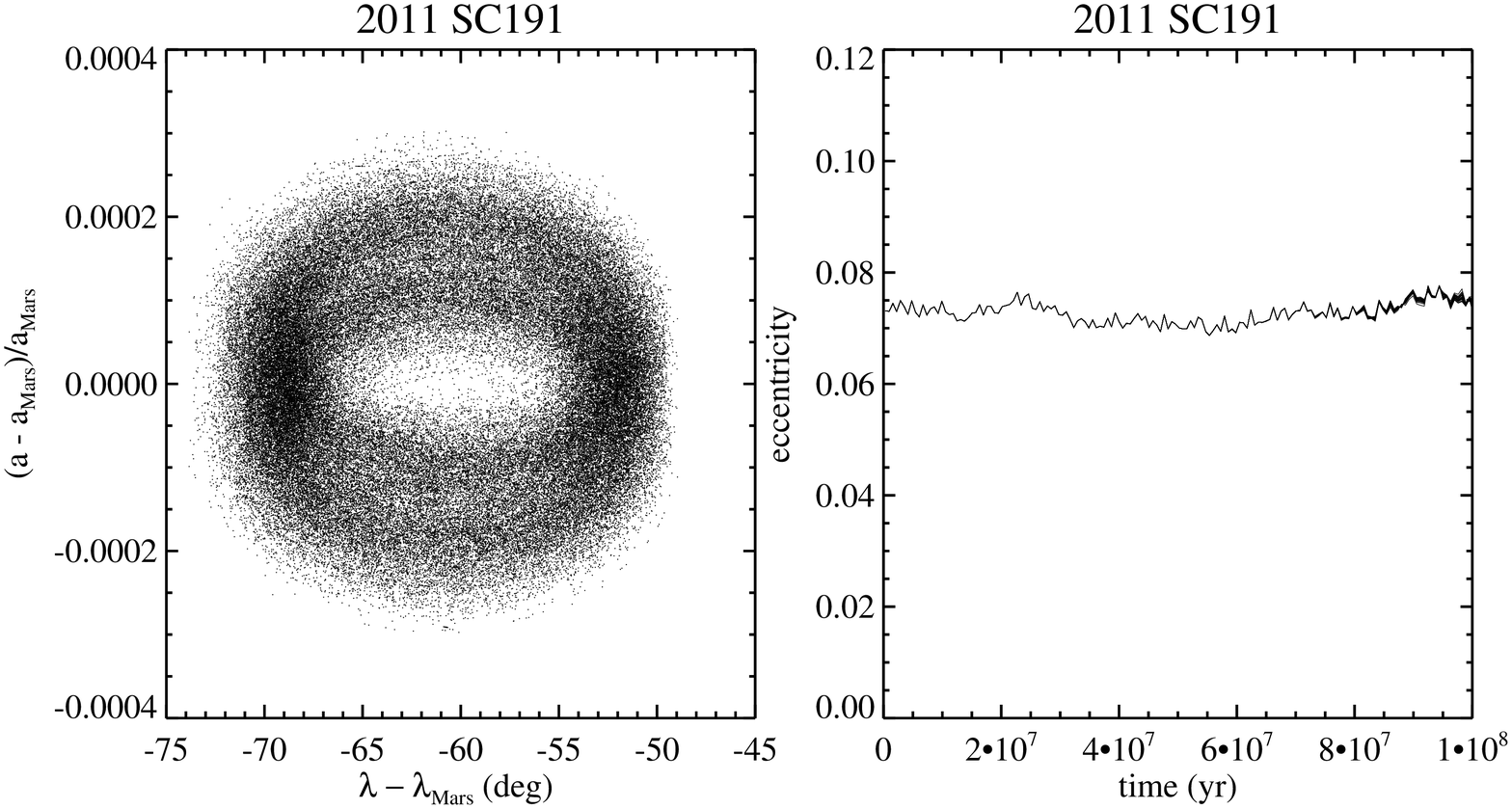}
\caption[The dynamical evolution of 404 clones (101 clones per object) in our $10^{8}$ yr integrations of the fully conservative system. 
Left panels: Longitude relative to Mars $\lambda - \lambda_{\rm Mars}$ vs relative normalised semimajor axis $\left( a - a_{\rm Mars}\right)/a_{\rm Mars}$.
Only one clone in every ten is shown for clarity. Right panels: 64-point running means of the osculating eccentricity vs time. 
Only one clone in five is shown for consistency with Fig~\ref{fig:1e8}.]{}
\label{fig:1e8noyark}
\end{figure}

\begin{figure}
\vspace{-3cm}
\centering
\includegraphics[height=4cm,angle=00]{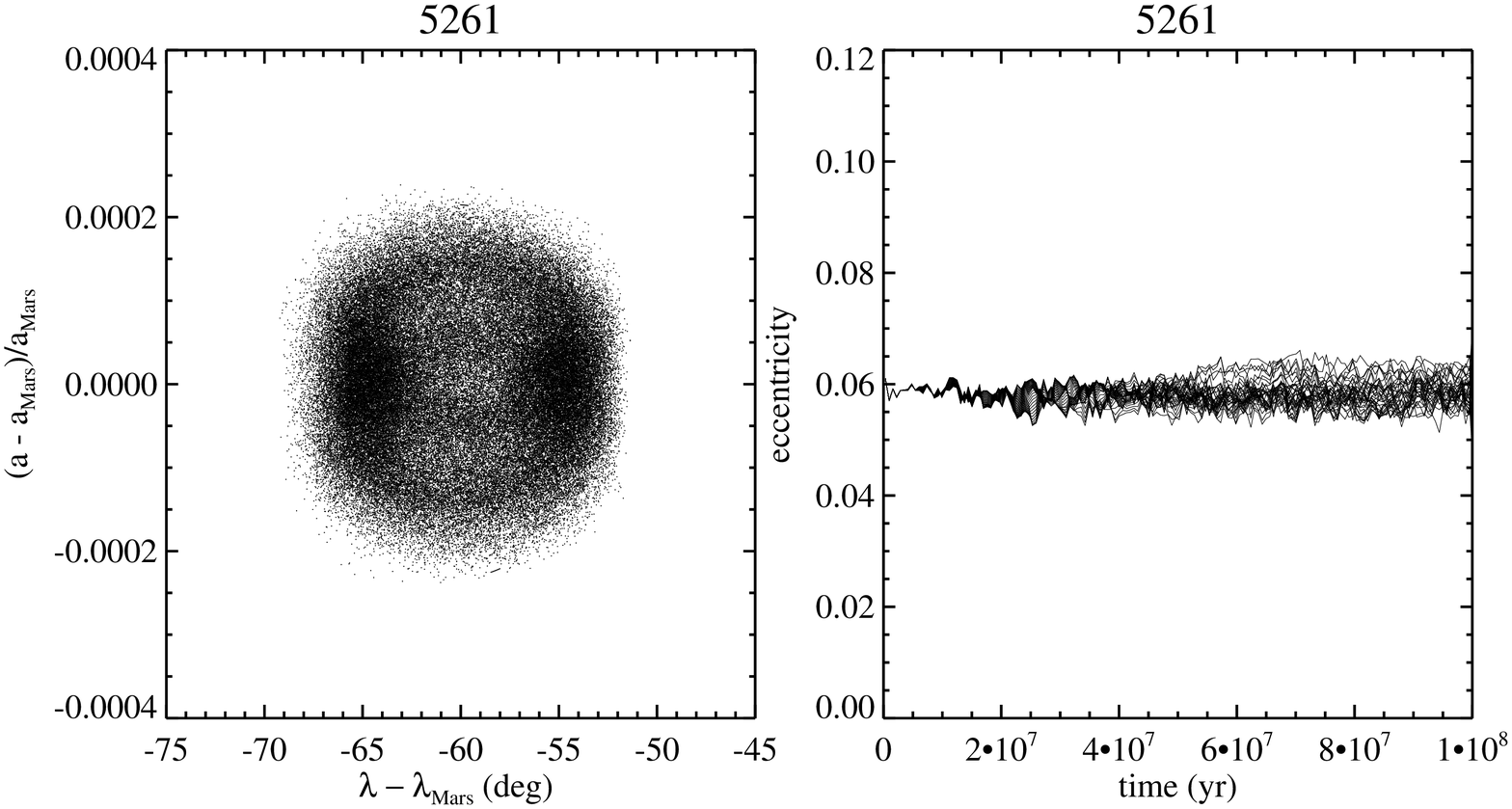}\\
\includegraphics[height=4cm,angle=00]{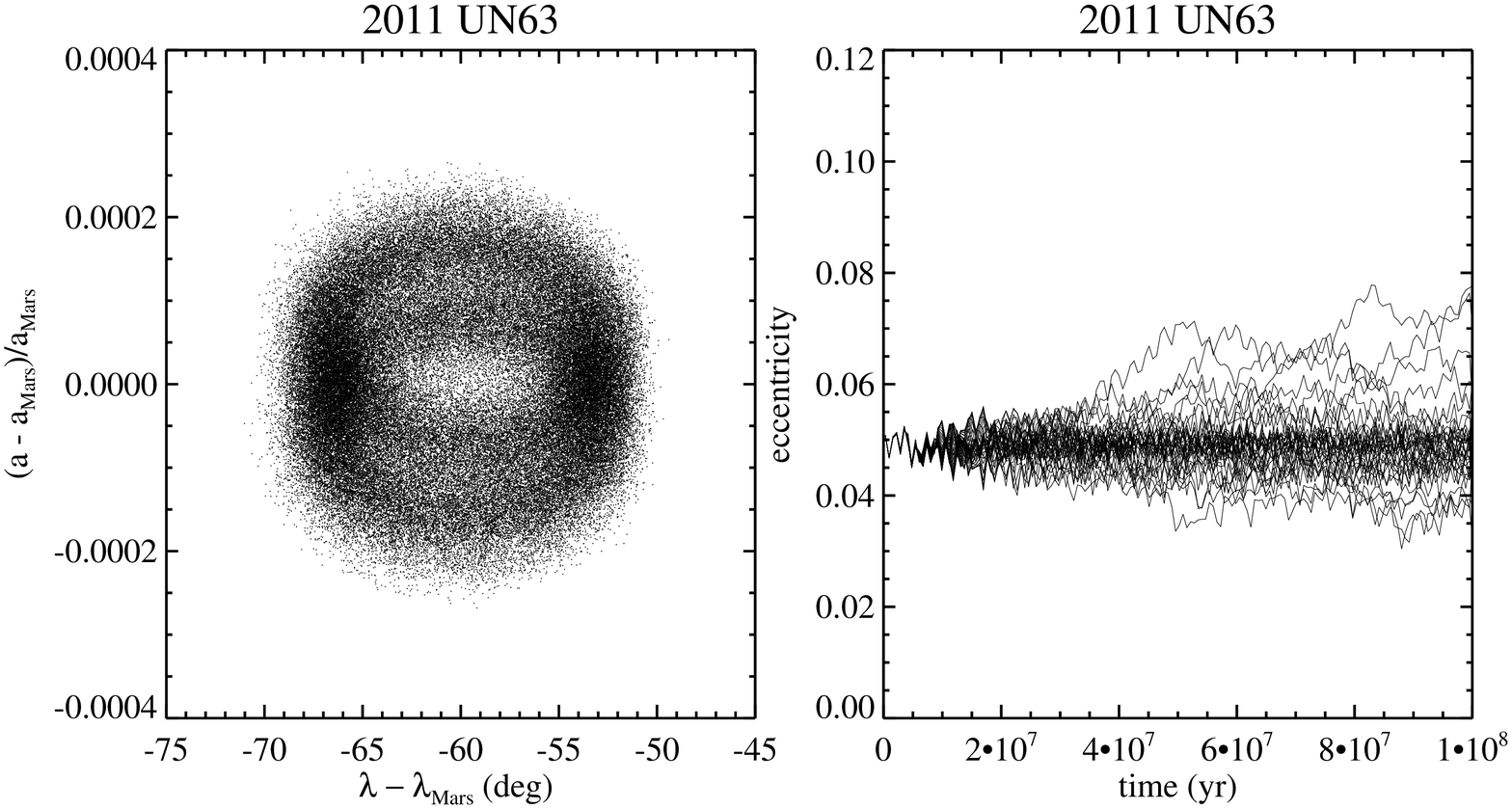}\\
\includegraphics[height=4cm,angle=00]{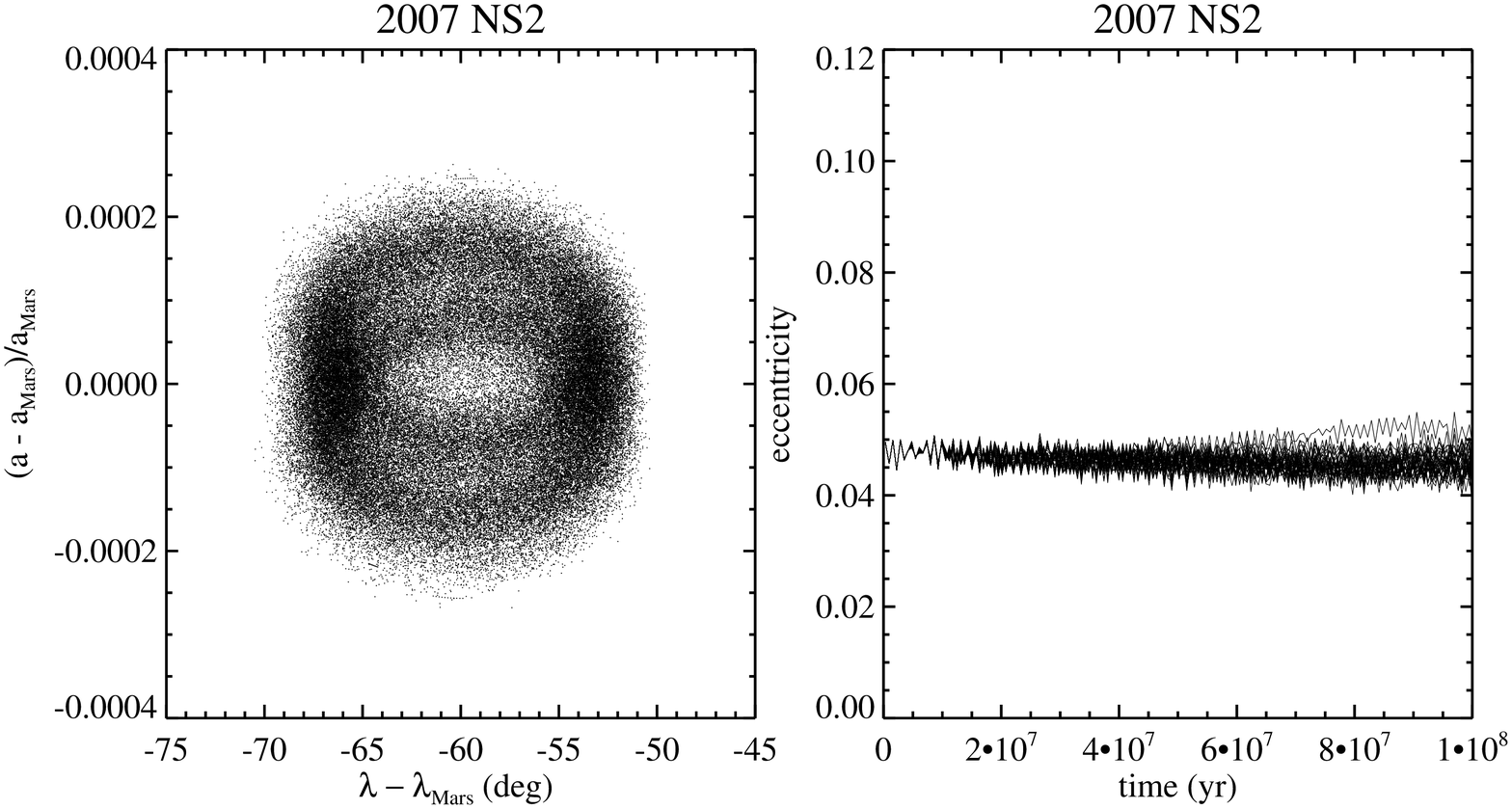}\\
\includegraphics[height=4cm,angle=00]{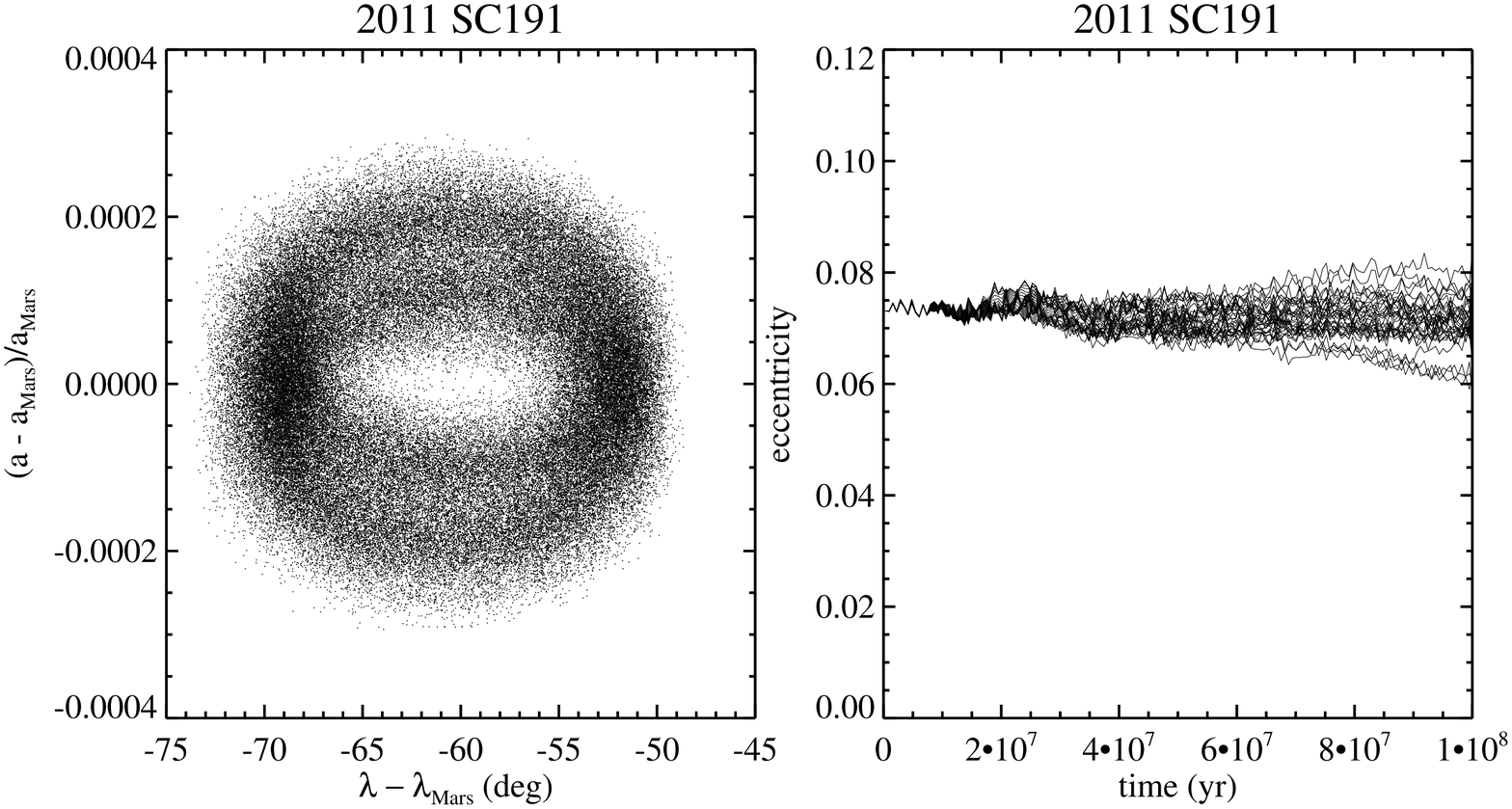}
\caption[As for Fig.~\ref{fig:1e8noyark} but including the model of the Yarkovsky force described in Section~\ref{supp:int}. Left: Only one clone in ten is shown for clarity. Right: Only one clone in five is shown for clarity.]{}
\label{fig:1e8}
\end{figure}

The results of the full $10^{8}$ yr simulations are presented in Figs~\ref{fig:1e8noyark} and \ref{fig:1e8}. For both batches, the relative longitude 
$\lambda_{r}= \lambda - \lambda_{\rm Mars}$ (left panels) for all 808 clones remains within $15^{\circ}$ of the $\mbox{L}_{5}$ equilibrium point. On the right panels we have plotted 64-point running means of the osculating eccentricity. In the first batch there is little or no dispersion or scatter between the clones' end states while significant dispersion is observed for the batch of clones that evolved under the Yarkovsky force. The higher scatter for clones of 2011 $\mbox{UN}_{63}$ and 2011 $\mbox{SC}_{191}$ in this latter batch is likely due to stronger Yarkovsky acceleration (as indicated by the higher values of $\dot{a}_{max}$ for these objects) although the larger uncertainty in their orbits compared to that of Eureka and 2007 $\mbox{NS}_{2}$ may also play a role. No evidence of gross instability or secular growth in the eccentricity is observed. On the basis of this evidence we conclude that all three new objects belong to the slow diffusing population (SD), able to remain as $\mbox{L}_{5}$ Mars Trojans for Gyr timescales.

\section{Estimating proper elements of Martian Trojans}
\label{supp:proper}
To establish how they relate to each other and the other known Trojans, particularly Eureka, we have calculated the proper elements of both new and previously known objects (Table~\ref{tab:sims}) by first carrying out a short ($7.3 \times 10^{5}$ yr) integration of their nominal orbits as described above with a 4d time step. The integration period was chosen so as to cover at least two periods of the $\nu_{5}$ mode ($\sim 3 \times 10^{5}$ yr).
The output was sampled with an interval of 128d to avoid aliasing of Mars' and the Earth's orbital frequencies. 
The output was processed with a programme utilising the Frequency Modified Fourier Transform algorithm \citep{SidlichovskyNesvorny1997}. Our technique for estimating the proper amplitude was somewhat different than that of SMT05. It is, essentially, the method of \citet{Milani1993} with two important differences: (i) we use the definitions of $d$ (the proper amplitude of the semimajor axis) and $D$ (the proper amplitude of the mean longitude) used in \citet{Marzari.et.al2003a,Marzari.et.al2003b} and SMT05, ie the ``full'' amplitude from minimum to maximum and twice the value listed in proper element catalogs and corresponding literature \citep{Milani1993,BeaugeRoig2001}, and (ii) the position of the equilibrium point is not fixed at $\pm 60^{\circ}$ but is taken directly from the harmonic analysis. Our estimate of $D$ for Eureka ($\sim 11.5^{\circ}$)
slightly underestimates the value ($15^{\circ}$) overplotted in Fig.~3 of SMT05. 
However, it is in good agreement with the value used in their Fig.~11. 
Uncertainties are estimated as the differences between the output of this procedure
and the results of carrying out the FMFT analysis on the first half of the time series only. As an additional test of the robustness of our 
$e_{p}$ and $I_{p}$ proper element estimates, we calculated 64-point running means of $e$ and $I$ for each clone of the three new objects and Eureka at the end of the 100 Myr 
integrations that included the Yarkovsky effect and determine lower and upper bounds after removing the five highest and lowest values (10\% of the sample). These bracket the proper element estimates obtained from harmonic analysis of the short-term integrations (Table~\ref{tab:sims}). In addition, our values of $d$ and $D$, estimated independently of each other, 
satisfy the approximate relationship $d \simeq \sqrt{3 \mu} a D$ where $\mu$ and $a$ are the Mars/Sun mass ratio ($\sim 3 \times 10^{-7}$) and semimajor axis ($1.52$ AU) of Mars respectively \citep{Erdi1988}. The relatively large uncertainty for $e_{p}$ in the case of 1998 $\mbox{VF}_{31}$ is probably due to its proximity to the $\nu_{5}$ secular resonance.

All four objects are contained within the domain $10^{\circ} < D < 20^{\circ}$, $0.04 < e_{p} < 0.075$ and ${18}^{\circ}.5< I_{p} < {22}^{\circ}.5$ (Table~\ref{tab:sims}). 
The corresponding dimensions of  the stable island found by SMT05 (illustrated in Fig.~\ref{fig:phase_space}) can be partitioned into 10 such bins in $D$ and 4 bins in each of $e_{p}$ and $I_{p}$.  
The phase space volume which contains Eureka and the three new Trojans is then $\sim 1/2 \times 10 \times 4 \times 4 = 100$ times smaller than the area of this stable island where the factor of
1/2 arises due to the roughly triangular projection of the latter in $e_{p} - I_{p}$. 

 The new objects were discovered by different NEO surveys and at different 
years and were chosen for this study for the quality of their orbits. Assuming a uniform proper element distribution of long-lived Martian Trojans, it is not clear why their proper elements should be so near each other and so similar to Eureka's. 

\section{Is the clustering of the proper elements of Eureka and the new objects statistically significant?}
\label{supp:stats}
\begin{figure}
\vspace{-3cm}
\centering
\includegraphics[width=120mm,angle=00]{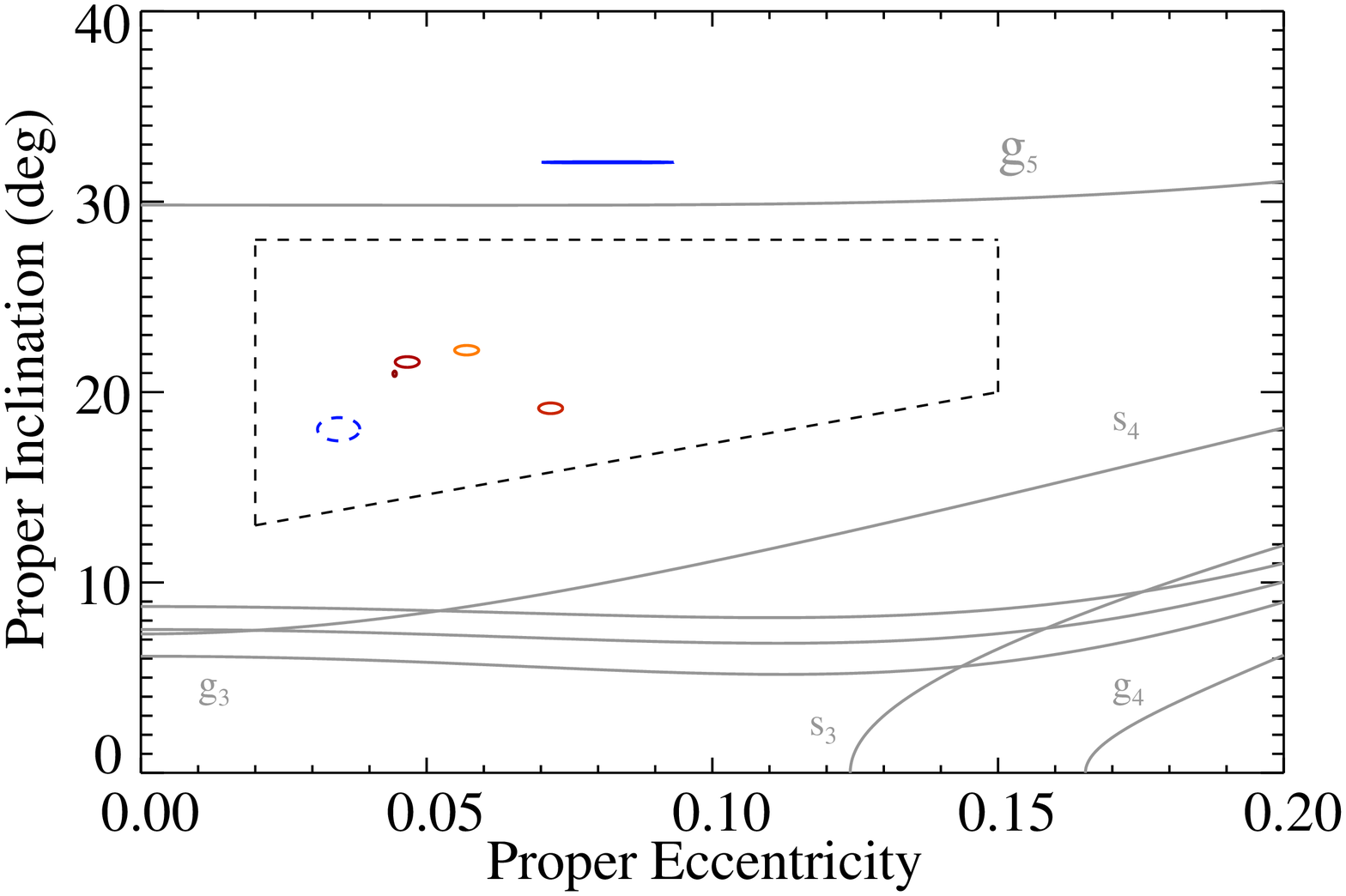}
\includegraphics[width=120mm,angle=00]{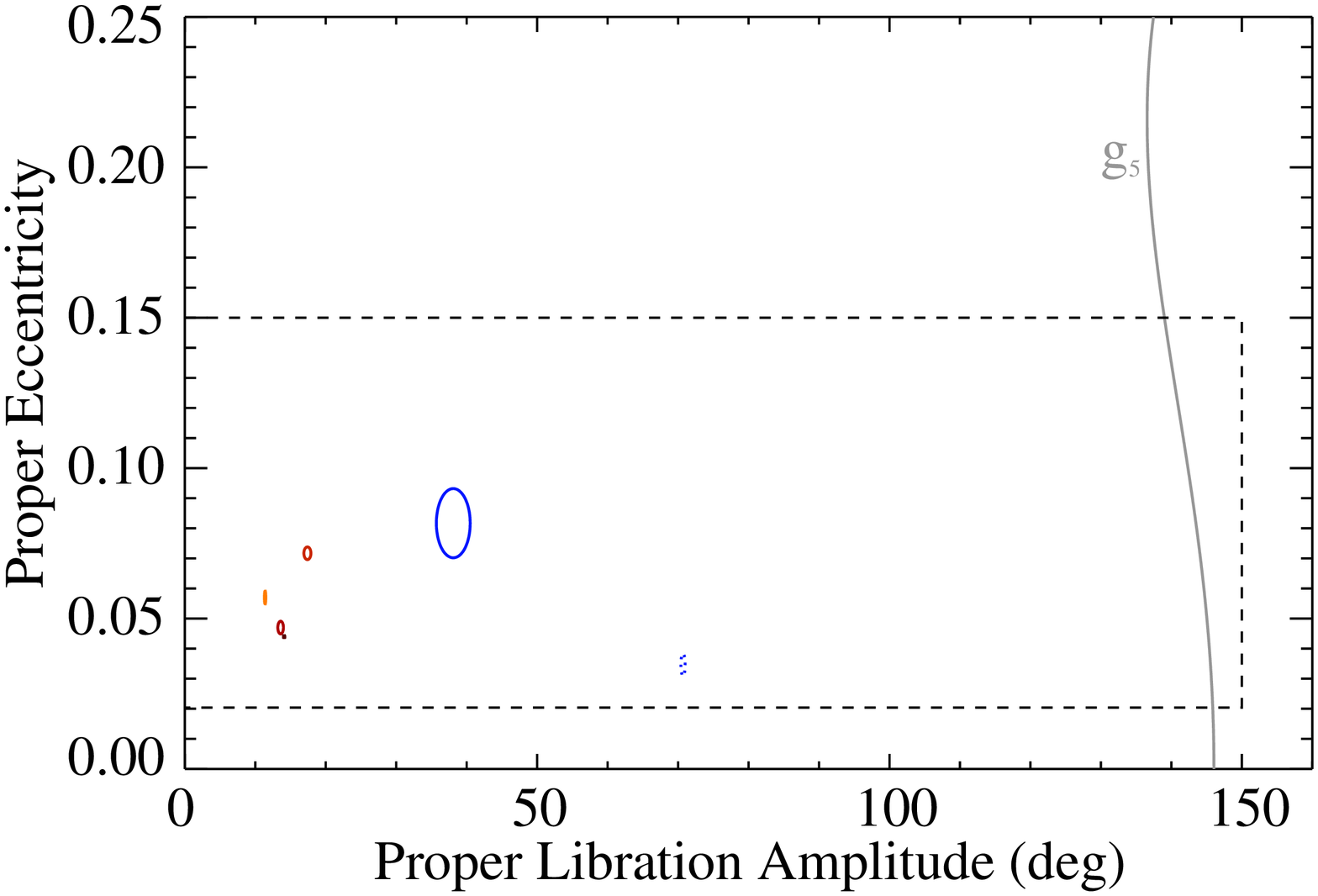}
\caption[Domains in proper element space shown by SMT05 to host the most stable Martian Trojans. These are enclosed within the dashed lines. Grey curves indicate the locations of secular resonances discussed in that work. The coloured ellipses indicate the location of the six Trojans and their semiaxes the uncertainties in their proper element estimates. Blue dashed curve: (121514) 1999 $\mbox{UJ}_{7}$, Blue curve: (101429) 1998 $\mbox{VF}_{31}$, Amber curve: (5261) Eureka, Brown curves: (301999) 2007 $\mbox{NS}_{2}$, 2011$\mbox{SC}_{191}$ and 2011$\mbox{UN}_{63}$. Top panel: Proper eccentricity ($e_{p}$) vs proper inclination ($I_{p}$). For secular resonance plotting, the proper amplitude of libration $D$ was fixed at $40^{\circ}$. Bottom panel: Proper amplitude of libration vs proper eccentricity. For secular resonance plotting, $I_{p}$ was fixed at $I_{p}=20^{\circ}$. (301999) 2007 $\mbox{NS}_{2}$ and 2011 $\mbox{UN}_{63}$ have been moved slightly farther apart for clarity.]{}
\label{fig:phase_space}
\end{figure}

Clustering tests used previously in this context \citep{Milani1993,BeaugeRoig2001,VokrouhlickyNesvorny2008} were designed to pick out object concentrations superimposed on a background population, but such a population is absent here.
To answer the question in our case, we estimate the probability that such a clustering would occur by chance
under the assumption of a uniform distribution of objects in $D$, $e_{p}$ and $I_{p}$. The validity of this assumption, which may depend on the formation scenario for these asteroids, will be discussed further in the Conclusions section. The projection of the domain containing long-lived Trojans (Fig.~1 of SMT05) on the  $e_{p} - I_{p}$ plane may be approximated by a quadrangle (Fig.~\ref{fig:phase_space}) defined by the points ($0.02, \mbox{ }13^{\circ}$), ($0.02, \mbox{ }28^{\circ}$), ($0.15, \mbox{ }28^{\circ}$) and ($0.15, \mbox{ }20^{\circ}$). The location of the $\nu_{5}$ resonance at $D \sim 140^{\circ}$ indicates that the region of stability may not extend as far as $D=150^{\circ}$. We proceed as follows: we assume that (a)
Martian Trojans can occupy a range of at least $100^{\circ}$ in $D$ (bottom panel of Fig.~\ref{fig:phase_space}) and (b) that their distributions in $e_{p}-I_{p}$ and in $D$ are uncorrelated.
We then partition the $D$ domain alone into 10 cells of $10^{\circ}$ width each. 

The problem of distributing $k$ consecutively-discovered Trojans into $n$ cells 
is equivalent  to seeking the integer, non-negative solutions of the equation  

$$ \sum_{i=1}^{n} x_{i} = k \mbox{.}$$

The number of discrete solutions (call it $C(n, k, 0)$) is 

$$ \left(\begin{array}{c} n + k - 1 \\  k \end{array} \right) \mbox{.}$$

If we impose the ``clustering'' constraint that $x_{i} \geq m$ for at least one $i$, the number of solutions $C(n, k, m)$ 
is found by assigning $x_{i}=m$ for one of the $i$ s and distributing the remaining $k-m$ units in the remaining $n-1$ cells.
Since there are $n$ choices for $i$ we have 

$$ C(n, k, m)  = n \left(\begin{array}{c} n + k - m - 2 \\  k - m \end{array} \right)$$ 

and the probability that a cluster of m (``m-cluster'') would occur by chance is 

$$ P(n, k, m) = C(n, k, m) / C(n, k, 0) \mbox{.}$$

For $n=10$, $k=5$ (all $\mbox{L}_{5}$ Trojans) and $m=4$ we get $C(10, 5, 0)=2002$ and $P(10, 5, 4)= 90/2002 \simeq 4.5 \times 10^{-2}$
or 1-in-22 odds of occurring by chance. 
Substituting $m=3$ in the above gives $450/2002 \simeq 0.22$ ie a 3-cluster has 2-in-9 odds of occurring by chance. 
The algorithm fails for $\left[k/m\right] > 1$ (square brackets denote the integer part) since different choices for $i$ no longer correspond to foreign sets of solutions. 
Now we increase the dimensionality of the model to include $e_{p}-I_{p}$ as well. The above probabilities are reduced
accordingly e.g. for a 3-cell equal-area partition of the trapezoidal domain (Fig.~\ref{fig:phase_space}, top panel) the reduction factors are $2/7$ for $m=4$ and $3/7$ for $m=3$ and
the respective probabilities become $1.3 \times 10^{-2}$ and $9.4 \times 10^{-2}$. Such a partition exists because all four Trojans
are contained within the box $18^{\circ}<I_{p}<23^{\circ}$, $0.04<e_{p}<0.075$ of surface area $< 1/3$ of that of the trapezoidal domain. The result is robust against adopting more conservative bounds for the trapezoidal
domain eg bounding $e_{p}$ at 0.11 instead of 0.15 and $I_{p}$ at $26^{\circ}$
instead of $28^{\circ}$.

Note on 2001 DH47 (see Section~\ref{supp:cands}): According to SMT05 it is an $\mbox{L}_{5}$ Trojan with $D \simeq 80^{\circ}$. If it is included in our analysis, then $k=6$ and the 
probability for $m=4$ evaluates to $P = 450 / 5005 \times 9/28 \simeq  2.9 \times 10^{-2}$. Thus the significance of the 4-cluster persists.

\section{Could the clustering of the proper elements of Eureka be due to observational bias?}
\label{supp:obs}

Observational searches for Martian Trojans are more likely to succeed within an area $\sim 30^{\circ}$ around the osculating mean longitude of each of the two triangular equilibrium points \citep{TabachnikEvans1999,TabachnikEvans2000b}. 
However, the new objects were discovered by different NEO surveys and at different years, chosen for this study for the quality of their orbits (Section~\ref{supp:cands}); osculating elements of Trojans do not relate to proper elements in a linear way \citep{BeaugeRoig2001};
two out of the three known long-lived Trojans have large $D$ ($\gtrsim 40^{\circ}$); and a recent search for $\mbox{L}_{5}$ Trojans yielded no new discoveries \citep{Trilling.et.al2006}. 
These are strong arguments in favour of (a) the orbital clustering of these four asteroids  - hereafter referred to as the Eureka cluster - being a real feature of 
the actual distribution of Martian Trojans and (b) the discovery of the new long-lived Trojans being the result of the gradually increasing sensitivity 
of small body surveys (ie higher limiting magnitude) rather than any observational bias towards Trojans with 
particular orbital properties. 

\section{Comparing the Trojan populations of Mars and Jupiter}
\label{supp:jupiter}
The Jovian Trojan clouds are well populated; the MPC lists 5161 Jupiter Trojans as of 21 June, 2012 (http://www.minorplanetcenter.net/iau/lists/ JupiterTrojans.html).
Their proper elements \citep{Milani1993,BeaugeRoig2001} are distributed throughout the long-term stable regions of phase space 
($e_{p} \lesssim 0.15$, $I_{p} \lesssim 35^{\circ}$ and $D \lesssim 40^{\circ}$) \citep{Marzari.et.al2003a}. Local concentrations in object density do exist, but unlike the case of the asteroid belt \citep{ZappalaCellino1992,MilaniKnezevic1994}, they are the exception rather than the rule \citep{BrozRozehnal2011}. 
To compare the Jovian and Martian cases, we partition the Jovian phase space into three-dimensional cells of width 
$\Delta e_{p} =0.05$, $\Delta I_{p} = 5^{\circ}$ and $\Delta D =10^{\circ}.$  We then read off currently available proper element catalogs 
one object at a time, placing it in the corresponding cell until one of the ten cells in the $\Delta D$ partition gains four (4) objects. 
The number of steps required to achieve this (the ``score'') is then a function of the $e_{p}$ and $I_{p}$ coordinates of the cell. 
The values of the proper amplitudes of libration read from the catalogs correspond to $D/2$ so they have been multiplied by 2 before use.
To assess the sensitivity of the results to the order that each Trojan is ``discovered'', we perform a cyclic shifting of the positions
of the first 100 Trojans (arguably the easiest ones to discover) with a randomly-chosen starting point and repeat the process 1000 times.
Trials with the first 20 and 50 Trojans yielded either the same or higher scores.
In this work we have considered the catalogs of semi-analytically-derived proper elements for 1702 numbered and multi-opposition Jovian Trojans by C.~Beaug\'{e} and F.~Roig (http://hamilton.dm.unipi.it/~astdys2/propsynth/petra.pro) and that of numerically derived proper elements of 1738 numbered and multi-opposition Trojans by Z.~Kne\v{z}evi\'{c} and A.~Milani (superseded in May 2012 by a new version containing 4030 entries; http://hamilton.dm.unipi.it/$\sim$astdys2/propsynth/tro.syn). The results are almost identical for the two catalogs so we resort to using the Beaug\'{e} and Roig catalog only. We have carried out the procedure in both the $\mbox{L}_{4}$ and $\mbox{L}_{5}$ clouds and find that the criterion is achieved earlier (ie at a lower score) for the $\mbox{L}_{4}$ cloud, consistent with the greater density of Trojans within the leading cloud \citep{BeaugeRoig2001}.

\begin{figure}
\vspace{-3cm}
\centering
\includegraphics[width=84mm,angle=0]{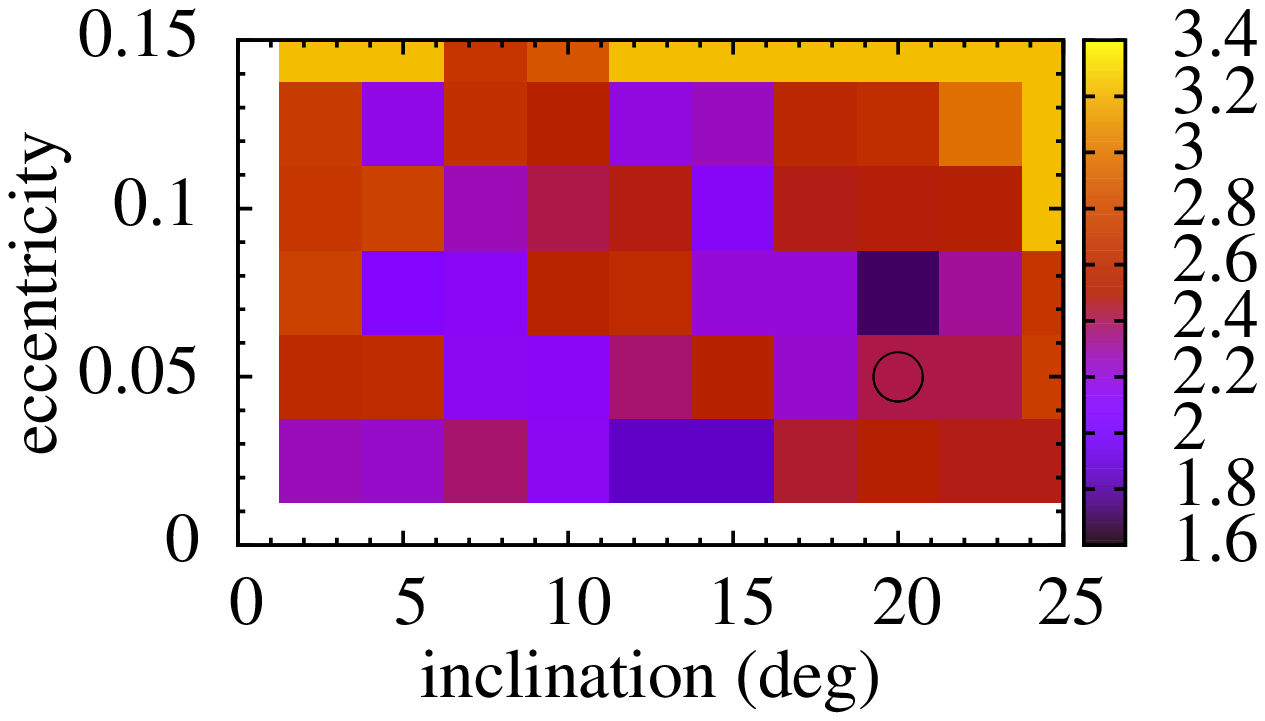}\includegraphics[width=84mm,angle=0]{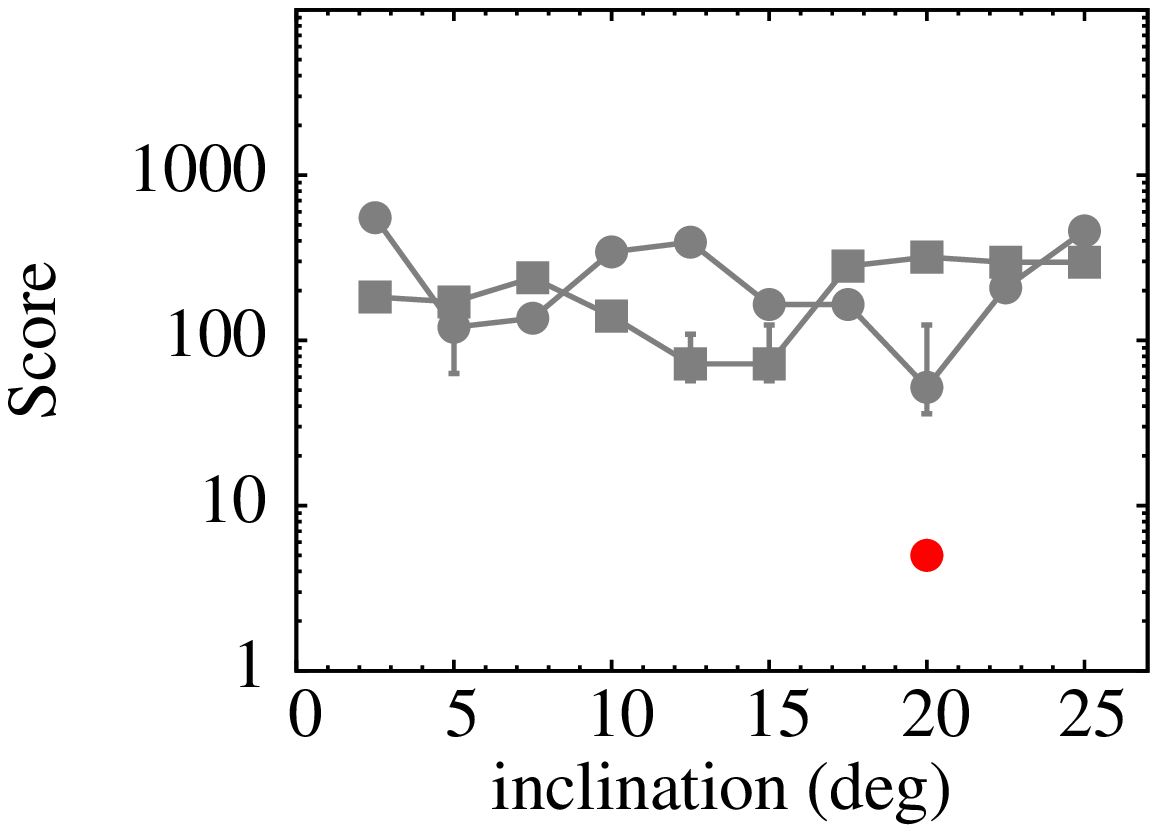}
\caption[Left panel:  Logarithmic density plot of the score achieved, by each resolution element in proper element space, 
of the population of known Jovian Trojans under the procedure described in the text. Each resolution element corresponds to objects in the 2-D domain $ \left( e_{p}-0.025, e_{p}+0.025 \right) \times \left( I_{p} - {2}^{\circ}.5 , I_{p} + {2}^{\circ}.5 \right) $. The open black circle indicates the location of the Mars Trojan cluster identified in this work. Right panel: Section of left panel for $e_{p} =0.025$ (filled grey boxes) and $e_{p}=0.075$ (filled grey circles). 
The red filled circle represents the result obtained in this work for Mars.]{}
\label{fig:contour}
\end{figure}

In the left panel of Fig.~\ref{fig:contour} we show the base 10 logarithm of the score achieved
for the ``nominal'' case, ie $\mbox{L}_{4}$ Trojans are considered in the order that they appear in the catalog. The minimum cell score
is 52 for the cell centered at $e_{p}=0.075$ and $I_{p}=20^{\circ}$. In the right panel we compare two one-dimensional cross-sections of the data, one for  $e_{p}=0.075$
and the other for $e_{p}=0.025$ with our result for Mars. Uncertainties are from randomly shifting the first 100 Trojan entries in the catalog. 
The minimum of 36 is achieved at the cell mentioned above. Hence, at least 30-40 and typically $\sim$100 $\mbox{L}_{4}$ Jovian Trojans need to 
be ``discovered'' before one finds a cluster with the same spread in proper elements as in the Martian case. 
From this exercise, we conclude that the clustering observed for the first few
Martian Trojans discovered cannot arise in the Jovian case and that the proper element distributions of the two populations are significantly different.

\section{Discussion: The origin of Martian Trojans}
The existence of a concentration of Trojans in the Martian $\mbox{L}_{5}$ region raises a number of questions 
regarding their origin and evolution.
Capture of stable Trojans by the terrestrial planets in the present solar system is unlikely \citep{Scholl.et.al2005,SchwarzDvorak2012}. 
In one origin model (A.~Morbidelli, reported in SMT05), these objects were captured as Trojans of Mars during the planet's random radial wandering in the early solar system  (cf Fig.~13 of that work). Alternatively, they could have been captured into the Trojan regions as a result of planetary mass growth \citep{FlemingHamilton2000}. Both scenarios imply deposition early in the solar system's history.
The three known Trojans are spectroscopically distinct.
Eureka belongs to the rare A/Sa taxonomic type \citep{Rivkin.et.al2003,DeMeo.et.al2009} not shared by 1998 $\mbox{VF}_{31}$ and 1999 $\mbox{UJ}_{7}$ \citep{Rivkin.et.al2007} and suggesting that Eureka was once part of a differentiated parent body. Thermal IR spectroscopy with the {\it Spitzer} spaceborne facility showed a high olivine content, consistent with either an oxidised chondritic or achondritic composition \citep{Lim.et.al2011}.
These objects' diverse taxonomic types are consistent with radial mixing of the main asteroid belt under a scenario requiring the early inward migration of Jupiter \citep{Walsh.et.al2011}. 
If this is what actually transpired, the deposition of Martian Trojans predates that of Jupiter Trojans, 
if the latter coincided with the so-called Late Heavy Bombardment as proposed recently \citep{Gomes.et.al2005,Morbidelli.et.al2005}. 

Orbital groupings of asteroids \citep{ZappalaCellino1992,VokrouhlickyNesvorny2008} can be produced either through collisions \citep{Michel.et.al2003} or rotational breakout \citep{Walsh.et.al2008} and fission \citep{Pravec.et.al2010} of a progenitor body.

Collisional lifetimes of Main Belt asteroids are a few times $10^{8}$ yr for objects of Eureka's size and a few times $10^{7}$ yr for smaller asteroids \citep{Obrien2004}. 
Collisions have been important in the evolution of Jovian Trojans \citep{Marzari.et.al1997}; Trojan-Trojan collision probabilities are higher \citep{Delloro.et.al1998} than those in the main belt due to the spatial confining action of the resonant libration. 
No studies directly relevant to Martian Trojan collisional evolution exist to-date but the two most important impactor populations will likely be 
(a) Mars Crossers (b) other Martian Trojans.  In the latter case, the efficiency of collisional comminution would depend on the (as yet unknown) 
population of small Trojan impactors. Due to the high inclination ($\sim 20^{\circ}$) of the Trojans and their proximity to the Sun, 
typical collision speeds with both populations would be more similar to those between MBAs and NEAs or NEAs and NEAs, $\gtrsim$ 10 km $\mbox{s}^{-1}$ \citep{Bottke.et.al1994}. For example, at the intersection point of two orbits with $a=1.524$ AU, $e=0$, $I=20^{\circ}$ and
$\Delta \Omega = 90^{\circ}$, the relative speed is $\sim 11$ km $\mbox{sec}^{-1}$. Collision probabilities may be enhanced relative to the Jovian case as the spatial azimuthal extent of the Trojan clouds scales as $a^{-1}$, the semimajor axis.

Studies of the collisional fragmentation of asteroids \citep{Asphaug.et.al1999,BenzAsphaug1999,Michel.et.al2003} apply mainly to lower velocity impacts and larger target bodies. 
 In recent hydrocode simulations \citep{Jutzi.et.al2009}, production of large fragments is enhanced 
at high impact velocities or km-sized targets while ejection velocities are a few m $\mbox{s}^{-1}$ at most 
for fragments $\gtrsim 1/10$ the size of the target. The resulting semimajor axis change of a fragment must be smaller than 
 the radial width of the Trojan region ($\sim a \sqrt{8 \mu/3}$ where $\mu$ is the planet's mass parameter and
 $a$ its orbital semimajor axis \citep{MurrayDermott1999}). The corresponding change in along-track velocity magnitude for an object in a circular, planar orbit 
 about the Sun is
 
 \begin{equation} \label{eq:dv}
 \Delta \mathrm{v} = \frac{1}{2} n  \sqrt{8 \mu /3} a
 \end{equation}
 
 where $n$ is the planet's orbital mean motion. Equation~\ref{eq:dv} evaluates to $\sim10$ m $\mbox{s}^{-1}$ for Mars and $\sim350$ m $\mbox{s}^{-1}$ for Jupiter. 
Fragment size is inversely proportional to ejection velocity and largest fragment ejection velocities increase with target size \citep{Jutzi.et.al2009}.
 This has two important consequences: (i) unlike the case of Jupiter \citep{Marzari.et.al1997}, all but the largest collisional fragments of Martian Trojan progenitors would have sufficiently high velocity to directly escape the Trojan regions, resulting in the observed paucity of Eureka-sized objects, and (ii)
the Eureka cluster's Trojan precursor could not have been larger than a few tens of km across, such as those postulated in some models of early giant planet migration \citep{Walsh.et.al2011}. For a mass ratio between parent body and largest fragment of $\sim 0.5$ (i.e. near the catastrophic disruption threshold \citep{BenzAsphaug1999}) the number of collisions required to produce an object of Eureka's size ($r=600$ m) is $10 \log R + 2.2 $ where $R$ is the parent body's radius in km. Setting $R = 5-50$ km yields 9-19 collisions over the age of the solar system, implying 4-8 collisions over the past 2 Gyr that the cluster likely formed (see below). We note that NEA-NEA collisional lifetimes for objects 1-30 km across are $\lesssim 10^{9}$ yr \citep{Bottke.et.al1994} but those in the Martian Trojan clouds could be significantly shorter as discussed above.

Rotational fission \citep{JacobsonScheeres2011} has been claimed to be responsible for the so-called ``spin barrier'' for NEAs \citep{Pravec.et.al2007} and pairs of Main Belt asteroids
with unusually similar orbits \citep{VokrouhlickyNesvorny2008,Pravec.et.al2010}. Characteristics 
of the mechanism are a separation velocity comparable to the escape velocity, $\sim 1$ m $\mbox{s}^{-1}$ for a $1$ km progenitor, 
and mass ratios $\lesssim 0.2$, implying a size ratio $\lesssim 0.6$. The latter constraint is satisfied for all 
possible pairs among these four asteroids that include Eureka and pairs that include 2007 $\mbox{NS}_{2}$ and the two smaller objects. 
Surface escape velocity can be expressed as a function of the object's radius $r$: $\mathrm{v}_{\rm esc} = \sqrt{8/3 \pi G \rho r^{2}}$ where $G$ is the gravitational constant ($\simeq 6.67 \times 10^{-11}$ N $\mbox{m}^{2}$  $\mbox{kg}^{-2}$) and $\rho$ the bulk density. If the secondary fissions away from the primary at escape velocity \citep{Pravec.et.al2010}, $r$ must be $<8$ km for $\rho = 3000$ kg $\mbox{m}^{-3}$ to prevent escape from the Trojan region (see previous paragraph). The fundamental difference between this mechanism and the collisional one is that ejection speed is not a strong function of ejectum size; therefore, even small secondaries (mass ratio $\ll 0.2$) should survive as Trojans for a small enough primary.  
The only available estimate of Eureka's rotation period \citep[$> $ 4.8 hr;][]{Rivkin.et.al2003} is not near the ``spin barrier'' \citep{Pravec.et.al2007} but is consistent with the expected primary rotation rate for the mass ratios of revelance here \citep{JacobsonScheeres2011}. We note that phase coverage of Eureka's lightcurve is incomplete; additional photometry is desirable to confirm the rotation rate value. 

Both mechanisms predict low ($< 10$ m $\mbox{s}^{-1}$) separation velocities for the three new Trojans if they originated from Eureka or a progenitor of comparable size. Applying the orbit comparison metric by \citet{Milani1993}, given by the expression

\begin{equation}
 n a \sqrt{(1/4) \left(\delta d / a \right)^{2} + 2 \left(\delta e_{p}\right)^{2} + 2 \left(\delta \sin I_{p}\right)^2}
\end{equation}

we find that the magnitudes of mutual orbital differences vary from $350$ m $\mbox{s}^{-1}$ (2007 $\mbox{NS}_{2}$ - 2011 $\mbox{UN}_{63}$) 
 to 1800 m $\mbox{s}^{-1}$ (5261 - 2011 $\mbox{SC}_{191}$). Hence, some orbital diffusion must have taken place within the Eureka cluster to modify $e_{p}$ and $i_{p}$. This can be used to constrain the time of formation. 
 
The proper elements of long-lived dynamical clones of Eureka diffuse appreciably over the age of  the solar system (SMT05).  The proper eccentricity diffuses faster, reaching values up to 0.15 over 2 Gyr. 
During the same period, the proper libration amplitudes of the clones remain similar to, and slightly larger than, Eureka's. They spread over a range of $\sim 50^{\circ}$ over the next 2 Gyr. 
The objects investigated here have $0.04<e_{p}<0.075$ and $10^{\circ}<D<20^{\circ}$ and their libration amplitudes are slightly larger than Eureka's. Consequently, if these orbital differences are
due to the action of chaotic diffusion alone (but see paragraph below), the event or events  that produced them have taken place less than 2 Gyr ago. 
This is consistent with a formation relatively late in the precursor's collisional history (ie a parent body originally tens of km across but close to Eureka's present size 
when the cluster was formed) in order to keep ejection velocities low. 

An additional complicating factor may be the action of the Yarkovsky effect, which broadens the proper element distribution of asteroid families on timescales $\gtrsim 10^{8}$ yr \citep{Bottke.et.al2001,Nesvorny.et.al2002b}. It is thought to be an inefficient agent of Trojan orbit modification \citep{Moldovan.et.al2010,BrozRozehnal2011} but may be more important in Mars' case since $\dot{a}_{\rm Y} \sim a^{-2}$ for reasonable choices of asteroid physical parameters \cite[see][and references therein]{Bottke.et.al2006}.
Dissipative forces in general modify the linear stability properties of Trojan motion \citep{Murray1994} in a way unique to the force considered. SMT05 found that the Yarkovsky effect depletes the Martian Trojan regions of objects $\lesssim 5$ m in size over the age of the solar system. In this work, we found that the Yarkovsky force affects the long-term evolution of the eccentricity. That such a size- and rotational state-dependent force has not dispersed the Eureka cluster beyond the point of recognition implies that either (a) it it inefficient, at least on the timescales of the cluster's age, or (b) it is similarly efficient on the members of the cluster. The latter possibility is unlikely given the disparate sizes of its members and their presumably different (but unknown) rotational axis orientations. We cannot, however, exclude the possibility that the high relative velocities ($>$ 300 m $\mbox{s}^{-1}$) between the orbits of cluster members are due, at least partly, to this effect. In that case, the dynamical age of the cluster would need to be revised downwards with respect to models dealing with chaotic diffusion alone. Although beyond the scope of the present study, follow-on work to quantify the long-term effects of the Yarkovsky force on Martian Trojans would be desirable.

An additional formation model is worthy of consideration here, that of the direct collisional capture of Trojans as proposed by \citet{ColomboFranklin1971} and recently explored
by \citet{Turrini.et.al2009} in the context of the origin of the irregular satellites of Saturn. In this scenario, a collision between two passing asteroids generates a spray of fragments and, 
among them, some have the right velocity to become Trojans. To explain the observations, such an event would have to produce objects within the limited
phase space volume occupied by Martian Trojans. This implies a fragment velocity field such 
that, in the first instance, the heliocentric semimajor axis dispersion is no more than a few times $10^{-3}$ AU. Indeed, it is not clear 
why such an event would not fill the stable island identified by SMT05 with fragments but instead produce the concentration 
of Martian Trojans reported in this paper. In conclusion, although such a scenario is possible, its {\it plausibility}
remains to be demonstrated.  

\section{Conclusions and Implications}

 This work has demonstrated the following: 
\begin{enumerate}
\item The existence of additional long-lived Trojans of Mars
\item A statistically significant concentration of Trojans around the largest known object of this class, 5261 Eureka
\item  A significant difference between the orbital distributions of Trojans of Mars and Jupiter
\end{enumerate}

A potential caveat on conclusions 2 and 3 is our assumption of an initially uniform distribution of Trojans. 
However, if the capture mechanism in the case of Mars produced an initial non-uniform distribution within the island of stability discussed in Section \ref{supp:proper}, the action of chaotic diffusion would have displaced these objects from their original locations and diluted any primordial features of their distribution over the age of the solar system.
Alternatively, one could argue that the rate of diffusion varies considerably for different position within this stable island but the results of SMT05, 
showing similar lifetimes for different long-lived Martian Trojans, do not support such a hypothesis.

Consequently, we have examined the following hypotheses for the existence of this cluster:
 
\begin{enumerate}
\item That it is the result of the collisional fragmentation of a progenitor body. This scenario is consistent with current understanding of 
asteroid collisional evolution and current knowledge of these Trojans' physical properties. No studies specifically applicable to Martian Trojans exist to-date.
\item That it is the result of the rotational fission of a progenitor body. This is also consistent with our understanding of the process and 
available physical information. In addition, it constrains the - largely unknown - rotational states of the asteroids.
\item That it is the result of direct collisional capture of asteroids in the Trojan region. We consider this a possible, but not yet a plausible, scenario 
pending a quantitative demonstration of its efficiency.     
\end{enumerate}

For the first two scenaria, the mutual orbital dispersion between members of this Trojan cluster implies an upper age limit of 2 Gyr.

New observational surveys will soon extend the known population of Martian and Terrestrial Trojans to $\lesssim 100$ m objects \citep{Todd.et.al2012a,Todd.et.al2012b}. The orbital distribution of new discoveries will constitute a test of the relative importance of the processes discussed above. If collisions are dominant, the paucity of km-sized Martian Trojans should be at least as apparent, if not more, at the smaller sizes. On the other hand, rotational fission will produce small asteroids that will remain within the Trojan clouds, shoring up the size distribution of the population at the small end. In effect, Trojan dynamics become a ``sieve'' to distinguish between different models of Martian Trojan production. Due to the unique geometric constraints of observing Terrestrial planet Trojans \citep{TabachnikEvans2000b}, the case of Mars is the most suitable to serve this function.
In any case, interpretation of the distribution of the newly discovered Martian Trojans, especially at the small end of the population, should be accompanied by (a) a re-examination of non-gravitational forces under the unique circumstances of Martian Trojan dynamics and (b) observational characterisation of key physical properties - size, albedo, taxonomic type and rotational state - of known objects and future discoveries.  
Finally, searches for genetic asteroid groups in the Terrestrial planet region are compromised by frequent planetary close approaches \citep{Fu.et.al2005,JacobsonScheeres2011,Schunova.et.al2012}. These, by definition, do not affect stable Trojans of the Terrestrial planets.  In this sense, our work highlights the Trojan regions of the inner solar system and, in particular, the Martian Trojan clouds as unique ``isolation chambers'' to identify asteroid production and family creation events close to the Earth and gain new insight on the physical and dynamical evolution of our solar system's small bodies.

\section*{Acknowledgements}
The author wishes to thank Prof.~Carl Murray for his insightful comments on earlier versions of the manuscript and
to acknowledge the SFI/HEA Irish Centre for High-End Computing (ICHEC) for the provision of computational facilities and support.
Astronomical research at the Armagh Observatory 
is funded by the Northern Ireland Department of Culture, Arts and Leisure 
(DCAL).
\bibliographystyle{elsarticle-harv}
\bibliography{icarus_christou_2013}
\clearpage
\protect
\listoffigures

\renewcommand{\baselinestretch}{1.0}
\protect

\newpage
\begin{table}[htb]
\centering
\caption[Orbital elements and physical properties of asteroids considered in this work.]{Orbital elements and physical properties of asteroids considered in this work.}
\begin{tabular}{lccccccc}
\noalign{\smallskip}
\hline \hline
\noalign{\smallskip}
Designation &   Epoch    &    $a$   &      &      $I$   &   &  $D$  &  \\
            &   (JD)   &     (AU)  &   $e$   &     (deg)  & $H$  &  (km) &  Opps \\  \hline
(5261) Eureka     &   2456000.5   &   1.523482  &  0.06483  &  20.28   &  16.01  & 1.28 &            13 \\
(311999) 2007 $\mbox{NS}_{2}$ &     "         &   1.523705  &  0.05399  &  18.62   &  17.75  & [0.67-2.1]  &    3 \\  
2011 $\mbox{UN}_{63}$         &     "         &   1.523709  &  0.06471  &  20.36   &  19.73  & [0.22-0.66]  &    2 \\
2011 $\mbox{SC}_{191}$        &     "         &   1.523811  &  0.04406  &  18.75   &  19.44  & [0.25-0.77]  &    3 \\
2011 $\mbox{SL}_{25}$         &   2455800.5   &   1.523888  &  0.11455  &  21.50   &  19.5   & [0.24-0.74]  &    1 \\
2011 $\mbox{SP}_{189}$        &   2455840.5   &   1.523279  &  0.04040  &  19.88   &  21.2   & [0.11-0.33]  &    1 \\
2011 $\mbox{UB}_{256}$        &   2455860.5   &   1.524049  &  0.07118  &  24.31   &  20.1   & [0.18-0.57]  &    1 \\
(101429) 1998 $\mbox{VF}_{31}$&   2456000.5   &   1.524213  &  0.01002  &  31.30   &  17.0  &  0.78  & 6 \\
(121514) 1999 $\mbox{UJ}_{7}$ &   2455800.5   &   1.524445  &  0.03915  &  16.74 & 16.9  &  [0.78-2.3]  &  6 \\ 
\noalign{\smallskip} \hline \hline \noalign{\smallskip} 
\multicolumn{8}{l}{\parbox{167mm}{The diameters of 5261 and 1998 $\mbox{VF}_{31}$ are from \citet{Trilling.et.al2007}. 
For the other objects we used the MPC's Absolute Magnitude to Diameter conversion tables to produce likely diameter ranges for albedos between 0.05 and 0.5.}}
\end{tabular} 
\label{tab:allobj}
\end{table} 
    
\clearpage
\begin{table}[htb]
\centering
\caption[Proper elements for all long-lived Martian Trojans from our simulations.]{Proper elements for all long-lived Martian Trojans from our simulations.}
\begin{tabular}{l@{}c@{ }ccccc@{ }c}
\noalign{\smallskip}
\hline \hline
\noalign{\smallskip}
                   &                 & $d$   & $D$        &           & $I_{p}$   &   &  $<I>$  \\
Designation        & Region & ($\times 10^{4}$AU) & (${}^{\circ}$) & $e_{p}$   & (${}^{\circ}$)  &    $<e>$ &  (${}^{\circ}$) \\\hline \noalign{\smallskip}
(5261) Eureka        &  $\mbox{L}_{5}$ &  ${2.84}^{\pm 0.02}$ &  ${11.4}^{\pm 0.1}$  &  ${0.0{\bf 570}}^{021}$  &   ${22.20}^{\pm 0.25}$    &  ${0.0{\bf 60}}^{05}$   &  ${22.35}^{\pm 0.25}$ \\
(311999)  2007 $\mbox{NS}_{2}$ &  $\mbox{L}_{5}$ &  ${3.54}^{\pm 0.01}$ &  ${13.9}^{\pm 0.1}$  &  ${0.0{\bf 444}}^{003}$  &   ${20.96}^{\pm 0.14}$    &   ${0.0{\bf 45}}^{04}$   &  ${21.0}^{\pm 0.4}$ \\
2011 $\mbox{UN}_{63}$          &  $\mbox{L}_{5}$ &   ${3.46}^{\pm 0.09}$ &  ${13.8}^{\pm 0.4}$  &  ${0.0{\bf 466}}^{021}$  &   ${21.58}^{\pm 0.28}$    &  ${0.0{\bf 525}}^{125}$   &  ${21.6}^{\pm 0.9}$  \\
2011 $\mbox{SC}_{191}$         &  $\mbox{L}_{5}$ &   ${4.42}^{\pm 0.15}$ &  ${17.4}^{\pm 0.5}$  &  ${0.0{\bf 717}}^{021}$  &   ${19.15}^{\pm 0.52}$     &   ${0.0{\bf 71}}^{06}$   &  ${19.15}^{\pm 0.65}$ \\
(101429) 1998 $\mbox{VF}_{31}$&  $\mbox{L}_{5}$ &  ${8.74}^{\pm 0.80}$   & ${38.1}^{\pm 2.4}$ & ${0.0{\bf 817}}^{115}$ &  ${32.07}^{\pm 0.03}$ & --  &  -- \\
(121514) 1999 $\mbox{UJ}_{7}$&  $\mbox{L}_{4}$ &   ${17.30}^{\pm 0.09}$   & ${70.7}^{\pm 0.3}$ & ${0.0{\bf 346}}^{037}$  & ${18.05}^{\pm 0.61}$ & -- &  -- \\
\noalign{\smallskip} \hline \hline  \noalign{\smallskip} 
\multicolumn{8}{l}{\parbox{167mm}{Columns 3, 4, 5 \& 6 give the proper amplitudes of the semimajor axis and mean longitude with respect to Mars, proper eccentricity and proper inclination from our FMFT analysis. The 7th and 8th columns give the range of values spanned by 64-point running means at the end of the $10^{8}$ yr integrations as explained in Section~\ref{supp:int}. 
Uncertainties are the superscripted values except for columns 5 and 7 where they replace the bolded digits.}}
\end{tabular} 
\label{tab:sims}
\end{table} 
 
\renewcommand{\baselinestretch}{2.0}

\clearpage

\clearpage

\clearpage

\clearpage

\clearpage

\end{document}